\documentclass[11pt]{article}
\usepackage{moriond,epsfig,amsmath,amssymb}

\newcommand{\beq}{\begin{equation}}
\newcommand{\eeq}{\end{equation}}
\newcommand{\bea}{\begin{eqnarray}}
\newcommand{\eea}{\end{eqnarray}}
\newcommand{\Fig}[1]{Fig.\,\ref{#1}}

\newcommand{\Eq}[1]{Eq.\,(\ref{#1})}

\newcommand{\f}{\frac}

\newcommand{\as}{\alpha_s}

\newcommand{\MW}{M_{\scriptscriptstyle W}}
\newcommand{\MZ}{M_{\scriptscriptstyle Z}}
\newcommand{\mc}{m_c}
\newcommand{\mb}{m_b}
\newcommand{\mt}{m_t}

\newcommand{\muh}{\mu_{\scriptscriptstyle W}}

\newcommand{\muc}{\mu_c}
\newcommand{\mub}{\mu_b}

\newcommand{\GF}{G_F}
\newcommand{\BR}{{\cal{B}}}
\newcommand{\GeV}{{\rm \ GeV}}

\newcommand{\MSbar}{\overline{\rm MS}}
\newcommand{\re}{{\rm Re}}
\newcommand{\im}{{\rm Im}}
\newcommand{\ord}{{\cal O}}

\newcommand{\sL}{{\scalebox{0.6}{$L$}}}

\newcommand{\Ktopinunu}{K^+ \to \pi^+ \nu \bar{\nu}}
\newcommand{\stodnunu}{s \to d \nu \bar{\nu}}

\newcommand{\KLtopinunu}{K_L \to \pi^0 \nu \bar{\nu}}

\newcommand{\Ktopienu}{K^+ \to \pi^0 e^+ \nu}
\newcommand{\Ktopinunus}{K \to \pi \nu \bar{\nu}}

\newcommand{\Pc}{P_c}
\newcommand{\dPuc}{\delta P_{c}}

\begin{document}
\rightline{ZU-TH 12/06; hep-ph/0605170}

\vspace*{4cm}
\title{THEORETICAL STATUS OF {\boldmath $\Ktopinunus$}
DECAYS\footnote{Talk given at the XLIst Rencontres de Moriond QCD and
High Energy Hadronic Interactions, La Thuile, Aosta Valley, Italy,
March 18--25, 2006.}}

\author{ ULRICH HAISCH }

\address{Institut f\"ur Theoretische Physik, Universit\"at
Z\"urich,\\ Winterthurerstrasse 190, CH-8057 Z\"urich, Switzerland}

\maketitle\abstracts{
We present a concise review of the theoretical status of rare
$\Ktopinunus$ decays in and beyond the standard model. Particular
attention is thereby devoted to the recent calculation of the
next-to-next-to-leading order corrections to the charm quark 
contribution of $\Ktopinunu$, which removes the last relevant
theoretical uncertainty from the $\Ktopinunus$ system.}

\section{\label{sec:introduction}Introduction}

The rare processes $\Ktopinunu$ and $\KLtopinunu$ play an outstanding
role in the field of flavor-changing-neutral-current transitions. The
main reason for this is their unmatched theoretical cleanness and
their large sensitivity to short-distance (SD) effects arising in the
standard model (SM) and its innumerable extensions. As they offer a
precise determination of the unitarity triangle (UT) \cite{UT}, a
comparison of the information obtained from the $\Ktopinunus$ system
with the one from $B$-decays provides a completely independent and
therefore critical test of the Cabibbo-Kobayashi-Maskawa (CKM)
mechanism. Even if these $K$- and $B$-physics predictions agree, the
$\Ktopinunus$ transitions will play an important part in discriminate
between different extensions of the SM \cite{Buras:2004uu}, as they
allow to probe effective scales of new physics operators of up to a
several TeV or even higher \cite{D'Ambrosio:2002ex} in a pristine
manner.     

\section{\label{sec:basic}Basic Properties of {\boldmath $\Ktopinunus$}}

The striking theoretical cleanness of the $\Ktopinunus$ decays is
linked to the fact that, within the SM, these processes are mediated
by electroweak (EW) amplitudes of $\ord (\GF^2)$, which exhibit a hard 
Glashow-Iliopoulos-Maiani (GIM) cancellation of the form    
\beq \label{eq:Aq}
{\cal A}_q (\stodnunu) \propto \lambda_q m_q^2 \propto \begin{cases}
\mt^2 ( \lambda^5 + i \lambda^5 ) \, , & \hspace{1mm} q = t \, , \\
\mc^2 ( \lambda + i \lambda^5 ) \, , & \hspace{1mm} q = c \, , \\
\Lambda_{\rm QCD}^2 \lambda \, , & \hspace{1mm} q = u \, .
\end{cases}   
\eeq
Here $\lambda_q = V^\ast_{qs} V_{qd}$ denotes the relevant CKM factors
and $\lambda = | V_{us} | = 0.225$ is the Cabibbo angle. This peculiar
property implies that the corresponding rates are SD dominated, while
long-distance (LD) effects are highly suppressed. A related important
feature, following from the EW structure of the SM amplitudes as well,
is that the $\Ktopinunus$ modes are governed by a single effective
operator, namely 
\beq \label{eq:Qv}
Q_\nu = \left (\bar s_\sL \gamma_\mu d_\sL \right ) \left ( \bar
{\nu}_\sL \gamma^\mu {\nu}_\sL \right ) \, ,  
\eeq
which consists of left-handed fermion fields only. The required
hadronic matrix elements of $Q_\nu$ can be extracted, including
isospin breaking corrections \cite{Marciano:1996wy}, directly from the
well measured leading semileptonic decay $\Ktopienu$. 

After summation over the three lepton families the SM branching ratios
for $\Ktopinunus$ can be written as
\cite{NLO,Isidori:2005xm,Isidori:2006qy}    
\beq \label{eq:BREQ}
\begin{split} 
\BR (\Ktopinunu) = ( 5.26 \pm 0.06 ) \left [ \left ( \f{\im
\lambda_t}{\lambda^5} X \right )^2 + \left ( \f{\re 
\lambda_t}{\lambda^5} X + \f{\re \lambda_c}{\lambda} \left (\Pc +
\dPuc \right ) \right )^2 \right ] \times 10^{-11} \, , \\ 
\BR (\KLtopinunu) = ( 2.29 \pm 0.03 ) \, \left ( \f{\im
\lambda_t}{\lambda^5} X \right )^2 \times 10^{-10} \, . \hspace{30mm} 
\end{split}
\eeq
The top quark contribution $X = 1.464 \pm 0.041$ \cite{NNLO}
accounts for $63 \%$ and almost $100 \%$ of the total rates. It is
known through next-to-leading order (NLO) \cite{X}, with a scale
uncertainty of slightly below $1 \%$. In $\Ktopinunu$, corrections
due to internal charm quarks and subleading effects, characterized by
$\Pc$ and $\dPuc$, amount to moderate $33 \%$ and a mere $4 \%$. Both
contributions are negligible in the case of the $\KLtopinunu$ decay,
which by virtue of \Eq{eq:Aq} is purely CP violating in the SM.         

\section{\label{sec:developments} Recent Developments in {\boldmath
$\Ktopinunu$}} 

\begin{figure}[!t]
\begin{center}
\begin{minipage}{0.45\textwidth}
\vspace{-7.5mm}
\hspace{6cm}
\scalebox{0.45}{\includegraphics{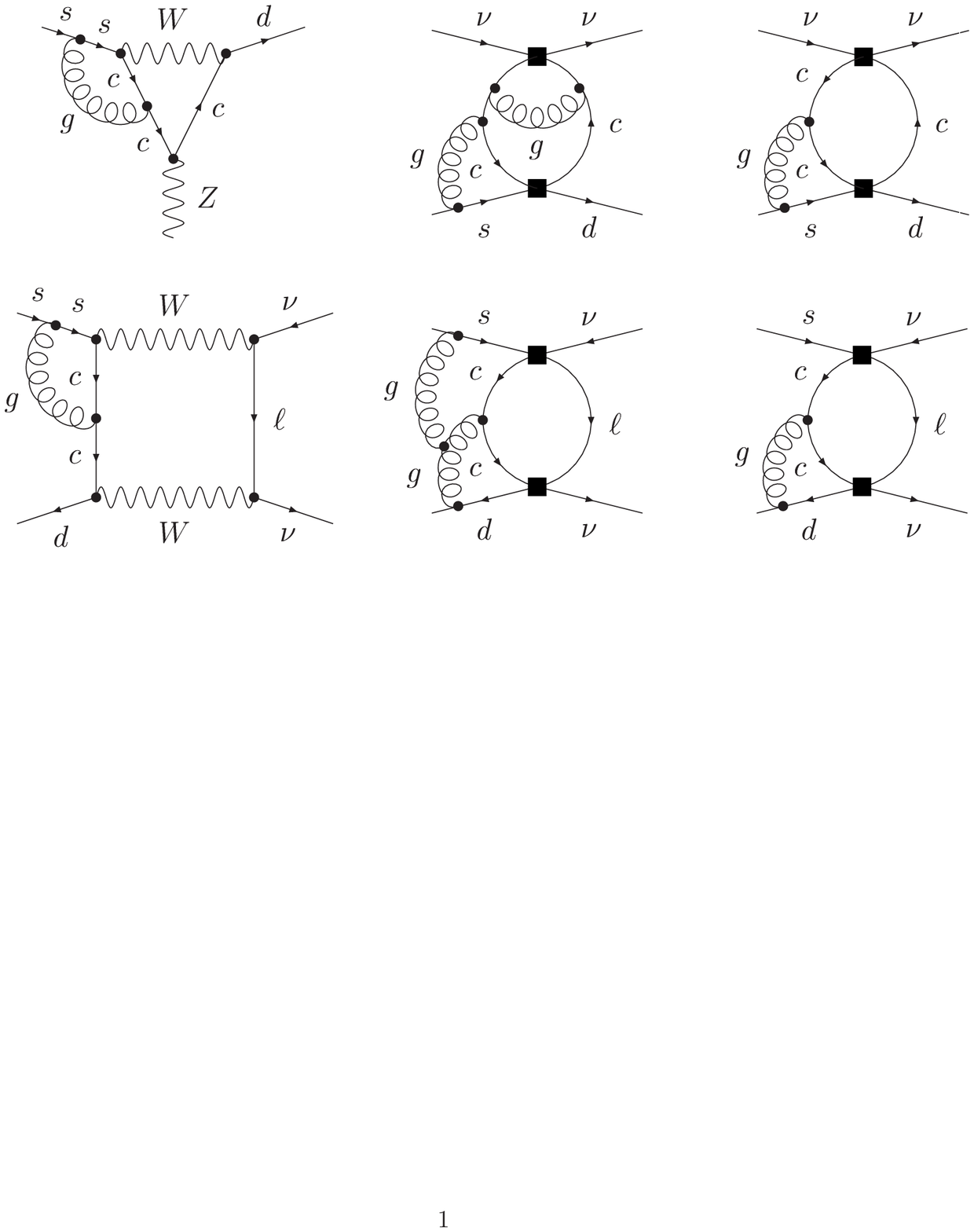}}
\end{minipage}
\begin{minipage}{0.45\textwidth}
\vspace{0mm}
\hspace{2.5mm}
\scalebox{0.725}{\includegraphics{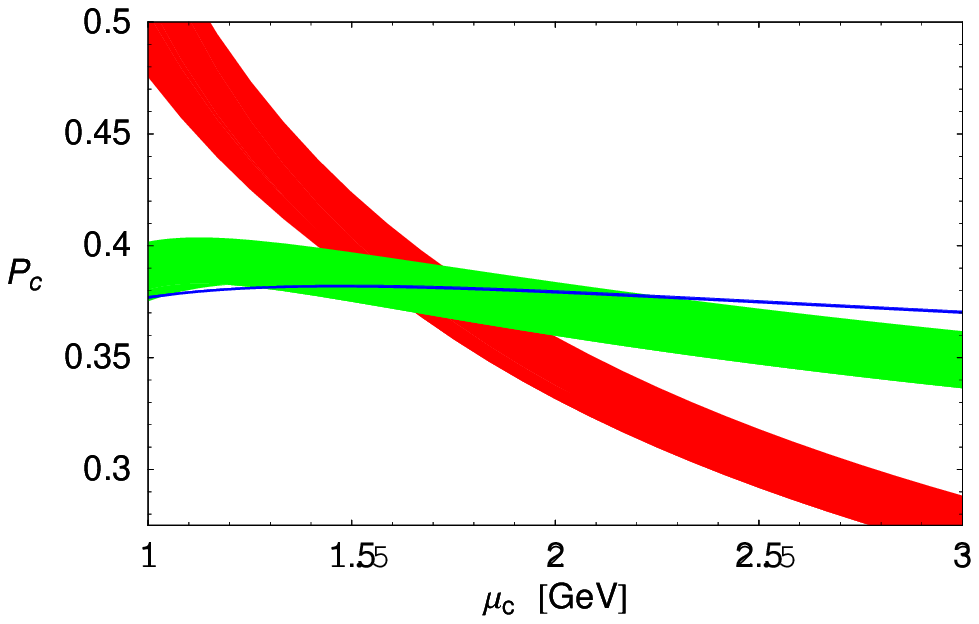}}
\end{minipage}
\end{center}
\caption{Left panel: Examples of diagrams appearing in the full SM
(left), describing the mixing of operators (center) and the matrix
elements (right) in the $Z$-penguin (top) and the electroweak box
(bottom) sector. Right panel: $\Pc$ as a function of $\muc$ at LO
(red), NLO (green), and NNLO (blue). The width of the bands reflects
the theoretical uncertainty due to higher order terms in $\as$ that
arise in the calculation of $\as (\muc)$ from $\as
(\MZ)$.~~~~~~~~~~~~~~~~~~}            
\label{fig:fig1}
\end{figure}

Two subleading effects, namely the SD contributions of dimension-eight
charm quark operators and genuine LD corrections due to up quark loops
have been calculated recently \cite{Isidori:2005xm}. Both
contributions can be effectively included by $\dPuc = 0.04 \pm 0.02$ 
in \Eq{eq:BREQ}. Numerically they lead to an enhancement of $\BR    
(\Ktopinunu)$ by about $7 \%$. The quoted residual error of $\dPuc$
can in principle be reduced by means of dedicated lattice QCD
computations \cite{Isidori:2005tv}.    

The main components of the state-of-the-art calculation of $\Pc$
\cite{NNLO}, are $i)$ the $\ord (\as^2)$ matching corrections
to the relevant Wilson coefficients arising at $\muh = \ord (\MW)$,
$ii)$ the $\ord (\as^3)$ anomalous dimensions describing the mixing of
the dimension-six and -eight operators, $iii)$ the $\ord (\as^2)$
threshold corrections to the Wilson coefficients originating at $\mub
= \ord (\mb)$, and $iv)$ the $\ord (\as^2)$ matrix elements of some of
the operators emerging at $\muc = \ord (\mc)$. To determine the
contributions of type $i)$, $iii)$, and $iv)$ two-loop
Green's functions in the full SM and in effective theories with five
or four flavors have to be computed including their finite
parts. Sample diagrams for steps $i)$ and $iv)$ are shown in the left
and right column of the left panel in \Fig{fig:fig1}. Contributions of
type $ii)$ are found by calculating the divergent pieces of three-loop
Green's functions with operator insertions. Two examples of Feynman
graphs with a double insertion of dimension-six operators are
displayed in the center column of the left panel in \Fig{fig:fig1}. 

The inclusion of the NNLO corrections removes essentially the entire
sensitivity of $\Pc$ on the unphysical scale $\muc$ and on higher
order terms in $\as$ that affect the evaluation of $\as (\muc)$ from
$\as (\MZ)$ which is sizeable at leading order (LO) and NLO. This is
explicated by the plot in the right panel of \Fig{fig:fig1} and by the
theoretical errors of the latest SM predictions \cite{NNLO}        
\beq \label{eq:Pc}
\Pc = \begin{cases} 0.369 \pm 0.036_{\rm theory} \pm 0.033_{\mc} \pm
0.009_{\as} \, , & \hspace{1mm} \text{NLO} \, , \\ 
0.375 \pm 0.009_{\rm theory} \pm 0.031_{\mc} \pm 0.009_{\as} \, , &
\hspace{1mm} 
\text{NNLO} \, .  
\end{cases} 
\eeq 
In obtaining these values the charm quark $\MSbar$ mass $\mc (\mc) =
(1.30 \pm 0.05) \! \GeV$ has been used. The residual error of 
$\Pc$ is now fully dominated by the parametric uncertainty from $\mc
(\mc)$. A better determination of $\mc (\mc)$ is thus an important
theoretical goal in connection with $\Ktopinunu$.         

Taking into account all the indirect constraints from the latest
global UT fit \cite{Charles:2004jd}, the updated SM predictions of the
two $\Ktopinunus$ rates read 
\beq \label{eq:BRSM}
\BR (\Ktopinunu) = \left ( 7.8 \pm 1.0 \right ) \times 10^{-11} \, ,
\hspace{5mm} \BR (\KLtopinunu) = \left ( 2.7 \pm 0.4 \right ) \times
10^{-11} \, .  
\eeq 
At present the errors from the CKM parameters veils the benefit of the
NNLO calculation of $\Pc$. However, given the expected improvement in
the extraction of the CKM elements and the foreseen theoretical
progress in the determination of $\mc (\mc)$, the isospin breaking
corrections, and in the case of $\BR (\Ktopinunu)$ of the LD
contributions, the allowed ranges of the SM predictions for $\BR
(\Ktopinunu)$ and $\BR (\KLtopinunu)$ should both reach the $5 \%$
level in the near future.       

\section{\label{sec:newphysics} New Physics in {\boldmath
$\Ktopinunus$}} 

An important virtue of the $\Ktopinunus$ modes is that new physics
(NP) contributions to their decay rates can be described, in almost
all beyond the SM scenarios, by just two parameters, namely the
magnitude of the SD function ${\cal X}$ and its complex phase
\beq \label{eq:XSD}
{\cal X} = |{\cal X}| \, e^{i \theta_{\cal X}} \, ,
\eeq
with $|{\cal X}| = X$ and $\theta_{\cal X} = 0$ in the SM. This
feature allows one to distinguish two classes of NP: $i)$ models
with minimal-flavor-violation (MFV), in which all flavor and CP
violation is governed by the structure of the SM Yukawa
interactions \cite{D'Ambrosio:2002ex}, implying that $\cal{X}$ is
real, and $ii)$ scenarios involving extra sources of flavor and CP
violation, in which $\cal{X}$ becomes complex in general.        

\begin{figure}[!t]
\begin{center}
\begin{minipage}{0.485\textwidth}
\scalebox{0.385}{\includegraphics{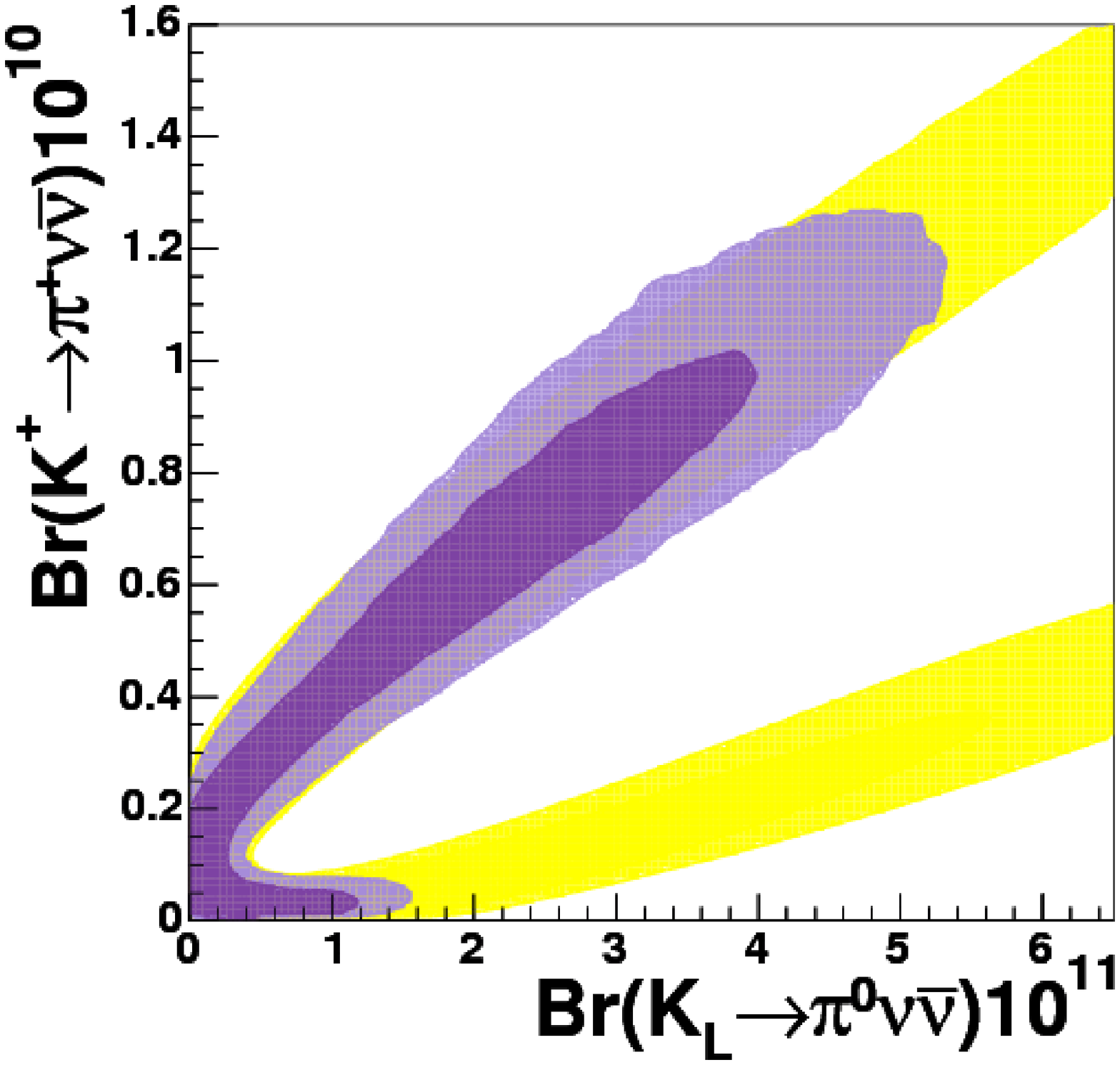}}
\end{minipage}
\hfill
\begin{minipage}{0.485\textwidth}
\scalebox{0.35}{\includegraphics{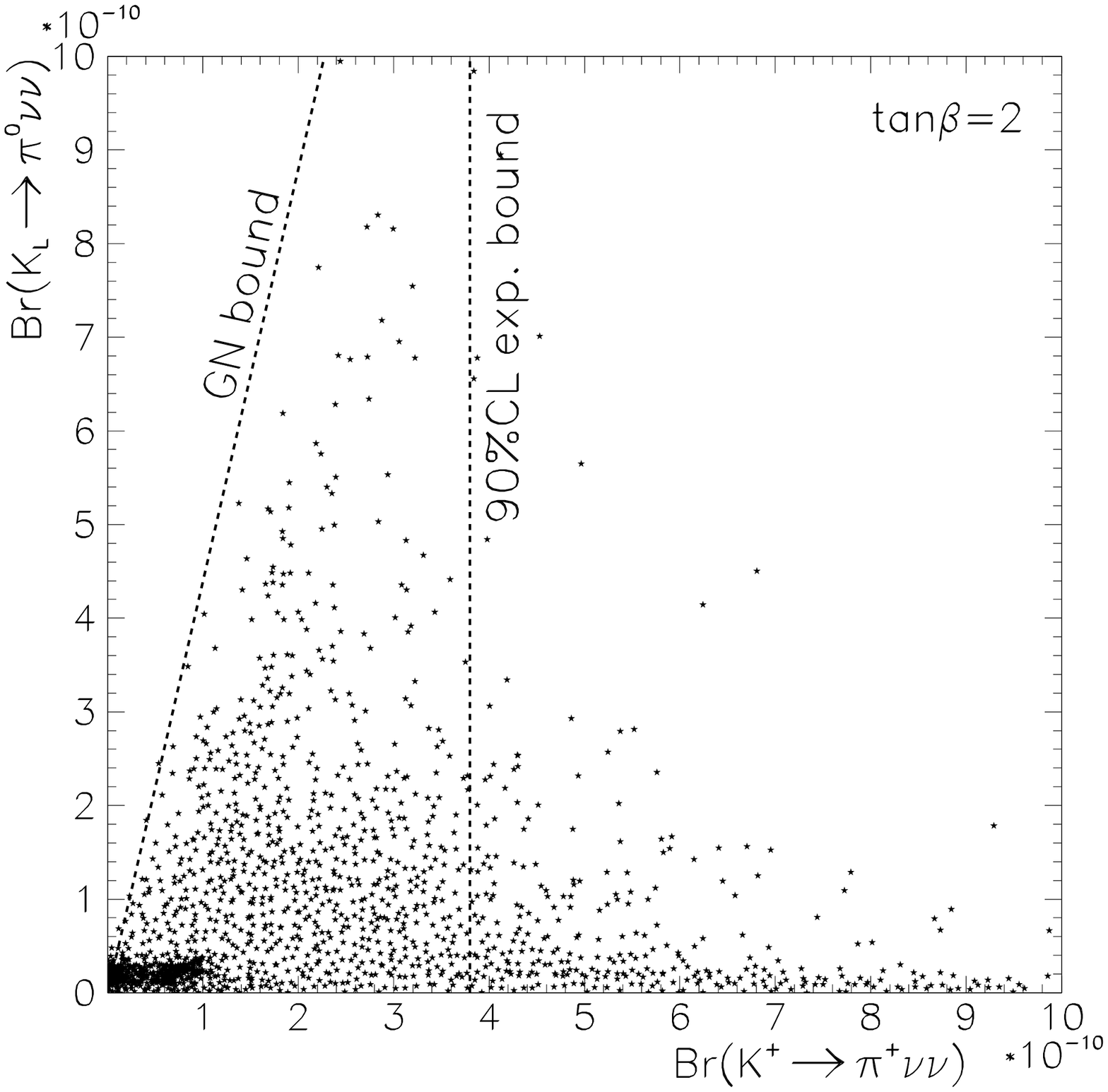}}
\end{minipage}
\end{center}
\caption{Left panel: Probability density function of $\BR
(\KLtopinunu)$ versus $\BR (\Ktopinunu)$ in MFV models. The dark
(light) eggplant shaped and colored areas correspond to the $68 \%$
($95 \%$) probability region while the yellow region represents the
allowed area without taking into account the experimental
information. Right panel: Distribution of $\BR (\Ktopinunu)$ versus
$\BR (\KLtopinunu)$ in the general MSSM for $\tan \beta =
2$. ~~~~~~~~~~~~~~~~~~~~~~~~~~~~~~~~~~~~~~~~~~~~~~~~~~~~~~~~~~~~~~}            
\label{fig:fig2}
\end{figure}

If one disregards the possibility of new local operators, MFV can be
formulated in terms of only 11 parameters: the real values of 7
universal master functions which describe the SD dynamics of a given
model and 4 CKM parameters that can be determined independent of loop
functions from the universal UT \cite{Buras:2003jf}. The specific 
formulation of MFV in terms of master functions allows one to study in
a transparent way correlations between different $K$- and
$B$-decays. An example of such a correlation is given in the left
panel in \Fig{fig:fig2} which shows the probability density function
of $\BR (\KLtopinunu)$ versus $\BR (\Ktopinunu)$. The corresponding
upper bounds for the branching ratios read \cite{Bobeth:2005ck} 
\beq \label{eq:MFVbound}
\BR (\Ktopinunu) < 11.9 \times 10^{-11} \hspace{2mm} (95 \% \, {\rm
CL}) \, , \hspace{5mm} \BR (\KLtopinunu) < 4.6 \times 10^{-11}
\hspace{2mm} (95 \% \, {\rm  CL}) \, ,  
\eeq  
which implies that in MFV models the $\Ktopinunus$ rates can be
enhanced by at most $20 \%$ and $25 \%$ relative to their SM
expectations. Very recently it has been shown \cite{Isidori:2006qy}
that these model-independent bounds can be saturated in the minimal
supersymmetric SM (MSSM) with MFV.           

A very different picture can emerge in models with new sources of
flavor and CP violation. Since now the GIM mechanism is no longer
active, large departures from the SM predictions are still possible
without violating any existing experimental constraint. Both
branching ratios can be as large as a few $10^{-10}$ with $\BR
(\KLtopinunu)$ often larger than $\BR (\Ktopinunu)$. In particular,
the Grossman-Nir (GN) bound \cite{Grossman:1997sk}     
\beq \label{eq:GNbound} 
\BR (\KLtopinunu) < \f{\tau (K_L)}{\tau (K^+)} \, \BR (\Ktopinunu) =
4.4 \, \BR (\Ktopinunu) \, ,   
\eeq 
following from isospin symmetry, can be saturated. This is illustrated 
by the plot in the right panel of \Fig{fig:fig2} which displays the
distribution of the allowed values of $\BR (\Ktopinunu)$ and $\BR
(\KLtopinunu)$ obtained by an adaptive scan involving 66 parameters 
of the general MSSM with conserved R-parity \cite{Buras:2004qb}. A
novel mechanism which could also lead to possible large deviations
from the SM predictions of the $\Ktopinunus$ rates in the context of
the MSSM with large $\tan \beta$ has been discussed recently
\cite{Isidori:2006jh}.   

\section{\label{sec:experiment} Experimental Situation}

Experimentally the $\Ktopinunus$ modes are in essence unexplored up to
now. The AGS E787 and E949 Collaborations at Brookhaven observed the
decay $\Ktopinunu$ finding three events \cite{KpEX}, while there is
only an upper limit on $\KLtopinunu$, improved recently by the E391a
experiment at KEK-PS \cite{KLEX}. The corresponding numbers read  
\beq \label{eq:BREX}
\BR (\Ktopinunu) = \left ( 14.7^{+13.0}_{-8.9} \right ) \times
10^{-11} \, , \hspace{5mm} \BR (\KLtopinunu) < 2.86 \times 10^{-7}
\hspace{2mm} (90 \% \, {\rm  CL}) \, .
\eeq 
Within theoretical, parametric and experimental uncertainties,
the observed value of $\BR (\Ktopinunu)$ is fully consistent with the
present SM prediction given in \Eq{eq:BRSM}.   

\section{\label{sec:conclusions}Conclusions}

An accurate measurement of $\BR (\Ktopinunu)$, either alone or
together with one of $\BR (\KLtopinunu)$, will provide an important
extraction of the CKM parameters that compared with the information
from $B$-decays will offer crucial tests of the CKM mechanism embedded
in the SM and all its minimal flavor violating extensions. The drastic
reduction of the theoretical uncertainty in $\Pc$ achieved by the
recent NNLO computation will play an important role in these efforts
and increases the power of the $\Ktopinunus$ system in the search for
new physics, in particular if $\BR (\Ktopinunu)$ will not differ much
from the SM prediction.    

\section*{\label{sec:acknowledgements}Acknowledgments}

I am grateful to A.~J.~Buras, M.~Gorbahn, and U.~Nierste for fruitful
collaboration. A big thank you to M.~Cacciari, G.~Dissertori,
A.~Gehrmann-De Ridder, M.~Grazzini, and G.~Zanderighi for not leaving
me lying bleeding in the snow. This work is supported in part by the
EU and the Schweizer Nationalfonds.  

\section*{\label{sec:references}References}

\end{document}